\documentclass[aps,prb,reprint,groupedaddress,superscriptaddress,showpacs]{revtex4-1}
\usepackage{color}
\usepackage{amssymb}
\usepackage{amsmath}
\usepackage{graphicx}
\usepackage{epstopdf}
\usepackage{amsmath}
\usepackage{amsthm}
\usepackage{latexsym}
\usepackage{graphicx}
\usepackage{dcolumn}
\usepackage{bm}
\usepackage{fancybox}

\begin{document}

\title{Polarization vortex domains induced by switching electric field  in
ferroelectric films with circular electrodes}
\author{Laurent Baudry}
\affiliation{Institute of Electronics, Microelectronics and Nanotechnology (IEMN)-DHS D\'epartment, UMR CNRS 8520, Universit\'e des Sciences et Technologies de Lille, 59652 Villeneuve d'Ascq Cedex, France}
\email{laurent.baudry@iemn.univ-lille1.fr}
\author{Ana\"{\i}s Sen\'{e}}
\affiliation{Laboratoire des Technologies Innovantes, Universit\'e de Picardie Jules Verne, Avenue des Facult\'{e}s - Le Bailly, 80025 AMIENS cedex , France}
\author{Igor A. Luk'yanchuk}
\affiliation{Laboratory of Condensed Matter Physics, University of Picardie Jules Verne,
Amiens, 80039, France}
\author{Laurent Lahoche}
\affiliation{Laboratoire des Technologies Innovantes, Universit\'e de Picardie Jules Verne, Avenue des Facult\'{e}s - Le Bailly, 80025 AMIENS cedex , France}
\author{James F Scott}
\affiliation{Department of Physics, Cavendish Laboratory,  University of Cambridge, Cambridge CB3 OHE, United Kingdom }
\date{\today}

\begin{abstract}
We describe ferroelectric thin films with circular electrodes and develop a
thermodynamic theory that explains exotic experimental results previously reported.
It is found to be especially useful for restricted geometries such as
microstructures for which boundary conditions are well known to play an
important role in ferroelectric properties. We have explored a new switching
mechanism, which consists of an inhomogeneous rotational motion of the
polarization and leads to an vortex state. The vortex appearance
exhibits characteristic properties of a first-order field-induced phase
transition with three critical electric fields and the possibility of hysteresis
behavior.
\end{abstract}

\maketitle

\preprint{APS/123-QED}

\affiliation{Institut d'Electronique de Micro\'{e}lectronique et de Nanotechnologie,
D\'{e}partement Hyperfr\'{e}quences et Semiconducteurs, UMR CNRS 8520,
Universit\'{e} des Sciences et Technologies de Lille,
Avenue Poincar\'{e}, BP 69, 59652 Villeneuve d'Ascq Cedex, France}


\affiliation{Laboratoire des Technologies Innovantes, Université de
Picardie Jules Verne, Avenue des Facult\'{e}s - Le Bailly, 80025 AMIENS
cedex , France}

\affiliation{Laboratory of Condensed Matter Physics, University of
Picardie Jules Verne, Amiens, 80039, France}

\affiliation{Laboratoire des Technologies Innovantes, Université de
Picardie Jules Verne, Avenue des Facult\'{e}s - Le Bailly, 80025 AMIENS
cedex , France}

\affiliation{Department of Physics, Cavendish Laboratory,  University of
Cambridge, Cambridge CB3 OHE, United Kingdom }



\section{INTRODUCTION}

The knowledge of spatial distribution of polarization in thin and ultra-thin
ferroelectric films is very important for their evaluation for device
applications. In fact, the finite-size ferroelectric samples exhibit
properties that are different from those for bulk materials \cite%
{DAWBER_2005}. Description of the nonuniform distribution of ferrelectric
polarizaration in frame of phase transition theory with suitable boundary
conditions has been successfully used to investigate the thin films
properties \cite{TILLEY_1984,TILLEY_1992}. It has been recently shown that
the domain textureis controlled by the surface layer
properties and related boundary conditions \cite{DEGUERVILLE_2005,LUKYANCHUK_2009}.
In term of mathematical formalism, changes in the boundary conditions for a partial
differential equation which are required for the determination of
variational problem solutions. 
The  modifications of physic properties  with respect to the bulk
induced by the presence of surfaces and/or interfaces can be investigated.
The corresponding solutions are non-trivial in that sense that they are different from those obtained for a
bulk material.

Unlike the vortex structures studied recently by Jia et
al.\cite{JIA_2011} or Balke et al. \cite{BALKE_2012} , these structures have
large radii (ca. 1 micron). The primary result is that there is a second
critical field $E_{\mathrm{c}}^{\mathrm{vortex}}$ a vortex nucleation field,
much lower than the usual coercive field $E_{c}$ for switching rectilinear
domains. This has device implications such as lower energy memory devices
for nonvolatile data storage. The polarization reversal process is a
succession of ortho-radial and radial stages. Simulation of the domain
texture evolution with time in two dimensions is found to be in very good
agreement with recent vortex structure dynamics reported by Gruverman et al.%
\cite{GRUVERMAN_2008}. Moreover, it is shown that the experimental doughnut
shape can exist in such a ferroelectric system but only as a metastable
state as described by Scott et al. \cite{SCOTT_2008}, which explains why it
disappears in hours.

\section{THERMODYNAMIC MODEL}

\begin{figure}[!h]
\centering
\includegraphics[angle=0,width=0.9\columnwidth]{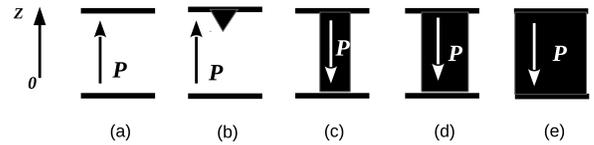}
\caption{Simplistic switching process stages.
(a) : homogeneous up-state ,  (b) : nucleation of a down-oriented region,
 (b) $\rightarrow$ (c): forward growth , (c) $\rightarrow$ (e): lateral growth ,
(e) homogeneous down-state.}
 \label{FIGA}
\end{figure}

The Landau theory of phase transition in the context of ferroelectric
material considers the switchable part of the polarization $\mathbf{P}$ as
the order parameter. Since the concept of an order parameter refers to the
description of an infinite media, the spontaneous polarization corresponds
to a global minimum of the free energy functional $F$. For thin films with
out-of-plane polarization along $z$-axis, the
experimentally observed switching mechanism from initial up-state toward 
a final down state is sketched in Fig.~\ref{FIGA}

Theoretically the direct switching mechanism as shown from (a) to (e) in Fig.~%
\ref{FIGA} inside a delimited region [i.e. without considering the
nucleation process (b)] can be described by using Landau approach in
the presence of an electric field.

Due to the anisotropy, one can consider the switching mechanism induced by
the electric field along the polarization axis $\mathbf{E}=E_{z}$ by using a
one dimensional approach  with the free energy functional  
$F=\int f\mathrm{d}V$ with $\mathbf{P}=P_{z}$ and

\begin{equation}
f(P_{z},E_{z})=\frac{\alpha }{2}P_{z}^{2}+\frac{\beta }{4}P_{z}^{4}+\frac{%
\kappa }{2}\left( \frac{\partial P_{z}}{\partial z}\right) ^{2}-E_{z}P_{z}
\label{EQ_LANDAU}
\end{equation}

\begin{figure}[!h]
\centering
\includegraphics[angle=0,width=0.5\columnwidth]{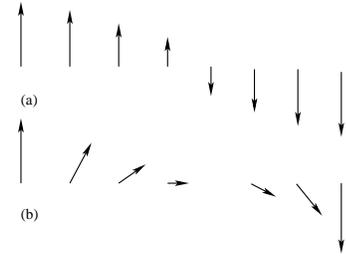}
\caption{Switching mechanisms : (a) longitudinal switching , (b) rotational
switching .}
\label{FIGB}
\end{figure}

In these conditions, the homegeneous switching mechanism with uniform $P_z$ uniform
occurs by mean of progressive change in the polarization modulus and an
abrupt change in sign when the field value reaches the thermodynamical
coercive field $E_c=-\frac{2\alpha}{3\sqrt{3}} \sqrt{\frac{-\alpha}{\beta}}$
[Fig.~\ref{FIGB}(a)]. This field value is many orders of magnitude higher
than the experimental coercive field \footnote{%
Discrepancy known as coercive field paradox}, the difference between these
two values being attributed to the crudeness of the nucleation process in
the thermodynamical description, so that the switching mechanism is
described as an homogeneous longitudinal phenomena 
(in reality ferroelectric switching always occurs via inhomogeneous nucleation).
 Some works have been devoted to extend this
description to inhomogeneous switching by including the gradient energy and
a surface term in the free energy \cite{BAUDRY_2005_2,BAUDRY_2007}, but
nevertheless without solving the coercive field paradox.

Possible solution of this paradox can be in the shown in [Fig.~\ref{FIGB}%
(b)] switching of the polarization by means of a rotational homogeneous
mechanism as it was proposed recently \cite{BAUDRY_2014}. In this case the polarization reversal occures without passing through
the state $P=0$, but rather by rotation of its direction. Such mechanism we
refer as " Iwata switching" can be relevant for the cubic perovkite-type
oxides in the bulk form, when the transition is described by the
three-component order parameter $\mathbf{P}=(P_{x},P_{y},P_{z})$, and the
free energy density $f$ has the following expression \cite%
{HU_1998,POTTER_2000}:

\begin{widetext}
\begin{eqnarray}
f(\mathbf{P},E_{z})&=&\frac{\alpha }{2}\left(P_{x}^{2}+P_{y}^{2}+P_{z}^{2}\right)
+\frac{\beta_1 }{4}\left(P_{x}^{4}+P_{y}^{4}+P_{z}^{4}\right)
 +\frac{\beta_2 }{2}\left(P_{x}^{2}P_{y}^{2}+P_{y}^{2}P_{z}^{2}+ P_{z}^{2}P_{x}\right)
 +\frac{1}{2}G_{11}\left[ (\partial_{x}P_{x})^{2}+(\partial_{y}P_{y})^{2}+(\partial _{z}P_{z})^{2}\right] 
 \nonumber \\
& & + G_{12}\left[ \partial_{x}P_{x}\partial_{y}P_{y}+\partial_{y}P_{y}\partial_{z}P_{z}+\partial_{x}P_{x}\partial _{z}P_{z}\right] 
+ \frac{1}{2}G_{44}[(\partial _{x}P_{y}+\partial _{y}P_{x})^{2}+(\partial
_{z}P_{y}+\partial _{y}P_{z})^{2}+(\partial _{x}P_{z}+\partial
_{z}P_{x})^{2}] 
+ \nonumber \\
& & \frac{1}{2}G_{44}^{\prime}[(\partial _{x}P_{y}-\partial _{y}P_{x})^{2} +
(\partial _{z}P_{y}-\partial _{y}P_{z})^{2}+(\partial
_{x}P_{z}-\partial_{z}P_{x})^{2}]-E_{z}P_{z} 
\label{EQ_IWATA}
\end{eqnarray}
\end{widetext}





However the potential problem of such scenario is the appearance of the
surface bound charges provided by the perpendicular polarization component
that is virtually emerged at the lateral sample during rotation. These
charges produce the energy-unfavorable depolarization field that finally
makes the homogeneous rotational switching inefficient in finite-size
samples.

The proposed mechanism however can be realized via the non-uniform rotation
of polarization vector inside the sample, when $\mathbf{P}$ stays parallel
to the surface and bound charges do not appear at the boundary. In addition,
the distribution of polarization inside the sample should satisfy the
condition $\mathrm{div}\mathbf{P}=0$ to not cause the internal bound charges
and related depolarization field.

\begin{figure}[!h]
\centering
\includegraphics[angle=0,width=0.5\columnwidth]{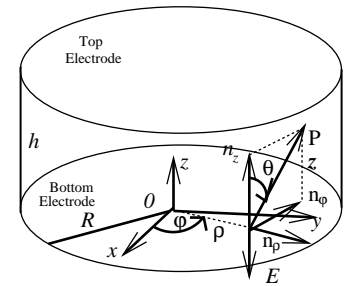}
\caption{Sketch of the structure studied.}
\label{FIGC}
\end{figure}

In present article we investigate the possibility of such nonuniform
switching, via the vortex formation mechanism as shown in [Fig.~\ref{FIGB}%
(b)]. To specify the geometry of the process we present the polarization
vector in cylindrical coordinates $\mathbf{P}=(P_{\rho },P_{\varphi },P_{z})$
[Fig.~\ref{FIGC}].

From basic electrostatic considerations related to symmetry and invariance
properties we have $P_{\rho }=0$, $P_{\varphi }(\rho )$ and $P_{z}(\rho )$.
For the present case concerned with vortex formation, one can expect that it
is also possible to switch the polarization by mean of progressive change in
the angle $\theta $ ; we now investigate carefully such kind of possibility
(i.e. switching by polarization rotation inside an ortho-radial plan with $%
\mathrm{div}\mathbf{P}=0$). In these conditions the free energy $f$ given by
Eq.~(\ref{EQ_IWATA}) becomes


\begin{eqnarray}
f(\mathbf{P},E_{z})&=&\frac{\alpha }{2}P_{\varphi }^{2}+\frac{\beta _{1}+\beta
_{2}}{4}P_{\varphi }^{4}+\frac{\alpha }{2}P_{z}^{2}+\frac{\beta _{1}}{4}%
P_{z}^{4}+\frac{\beta _{2}}{2}P_{\varphi }^{2}P_{z}^{2} \nonumber\\ 
& & {}+G\left[ \left( \frac{\partial P_{z}}{\partial \rho }\right) ^{2}
+\left( \frac{1}{\rho }\frac{%
\partial (\rho P_{\varphi })}{\partial \rho }\right) ^{2}\right] -E_{z}P_{z},
\label{EQ_CYLIN}
\end{eqnarray}%

where $G=G_{44}+G_{44}^{\prime }$. 

Further simplification can be done close to the morphotropic phase boundary
where the anisotropic terms in Eq.~(\ref{EQ_CYLIN}) are negligible and the
rotational degrees of $\mathbf{P}$ are not coupled with its modulus. In this
case we can use the constant modulus Goldstone approximation and
parametrize the radial polarization distribution structure by the only
parameter, which is the inclination angle $\theta (\rho )$.
%
By introducing scalar variables $P$, $E$ related to the vectors components $%
P_{z}$, $P_{\varphi }$, $E_{z}$ by $P_{z}=P\cos \theta $, $P_{\varphi
}=P\sin \theta $ and $E_{z}=-E$ (with $E>0$).
We obtain the expression of the free energy with  $\theta =\theta (\rho )$:

\begin{eqnarray}
f(P,E)&=&\ GP^{2}\left[ \left( \frac{\partial \theta }{\partial \rho }\right)
^{2}+\frac{1}{\rho ^{2}}\sin ^{2}\theta +\frac{1}{\rho }\frac{\partial \sin
^{2}\theta }{\partial \rho }\right] \nonumber \\
& & {}-PE\cos \theta .
\end{eqnarray}

It is convinient to use the rescaled quantities $\tilde{f}=\frac{f}{%
-\alpha _{0}\zeta _{0}^{2}P^{2}}$ and $\tilde{E}=\frac{E}{-\alpha _{0}\zeta
_{0}^{2}P}$ where 
$\zeta _{0}=\sqrt{G/-\alpha _{0}}$ and $\alpha _{0}=\alpha \mid
_{T=0}=-aT_{c}<0$. The parameter $\alpha =a(T-T_{c})$ being analogous to the
coefficient of $P^{2}$ in the classical Landau free energy expansion (Eq.~(%
\ref{EQ_LANDAU})).

Finally we obtain the expression of the rescaled free energy functional $%
\tilde{F}$ of the ferroelectric cylinder in the presence of the applied field

\begin{equation}
\tilde{F}=\int_{0}^{2\pi}\int_{0}^{R}\int_{0}^{h} \tilde{f}(\rho,\theta)
\rho \mathrm{d} \rho \mathrm{d} \varphi \mathrm{d} z,
\end{equation}

with

\begin{equation}
\tilde{f}(\rho,\theta)=\left( \frac{\partial \theta }{\partial \rho}\right)
^{2} +\frac{1}{\rho^2} \sin^2 \theta +\tilde{E} \cos \theta +\frac{1}{\rho }%
\frac{\partial \sin ^{2}\theta }{\partial \rho }.
\end{equation}

We are interested in a thin film homogeneously up-polarized at the initial
state, i.e. with $\theta_i (\rho )=0$ , and would like to determine the possible
final states obtained after applying a down electric field by mean of
circular electrode with radius $R$. One can expect that there exist an
intuitive solution which consists in a homogeneously down polarization state
in the region where the field is applied (i.e. located under the circular
electrode). In other words this solution for final state corresponds to $%
\theta_f (\rho)=\pi $ on the interval $[0,R]$. This solution is also from
both mathematical and physical viewpoints a trivial solution when the
diameter of the electrode is infinite, because it corresponds to a global
minimum (polarization parallel to the field) for the free energy for an
infinite system. 
For the case of real systems with finite electrode diameter, the finite
character of the ferroelectric media would play an important role and the
interaction with the exterior will be crucial via boundary conditions.

We have to solve a variational problem to search for functions $\theta(\rho)$
which are extrema of the functional $\tilde{F}$ and first develop the
variation of the functional by carefully paying attention to the
contribution of boundary conditions. 


\begin{eqnarray}
\delta \tilde{F}&=&2\pi h\int_{0}^{R}g(\theta (\rho ,e))\delta \theta \mathrm{d}\rho 
+2\pi h\left[ \frac{\partial \theta }{\partial \rho }\delta \theta \right] _{0}^{R}  \nonumber \\
& &  +2\pi h\left[ \frac{1}{2}\sin 2\theta \delta \theta \right]
_{0}^{R},
\end{eqnarray}%

with%
\begin{equation}
g(\theta (\rho ,e))=-2\nabla _{\rho }^{2}\theta +\frac{1}{\rho ^{2}}\sin
2\theta -e\sin \theta .
\end{equation}

In order to find non trivial solutions we have assumed that the polarization
was pinned in the initial up-oriented polarization state $\rho =0$ and $\rho =R$, so that the value of $\theta (0)=0$  and $\theta (R)=0$  are fixed and independed on the field.


Finally variational calculation leads to the following state equation
expressed in terms of rescaled variables $\tilde{\rho}=\rho /R$ and $\tilde{e%
}=\tilde{E}R^{2}\geq 0$ :

\begin{equation}  \label{EQ_ETAT_NONLIN}
\frac{\partial^2 \theta}{\partial \tilde{\rho}^2} +{\frac{1 }{\tilde{\rho}}}
\frac{\partial \theta}{\partial \tilde{\rho}} -\frac{1}{2\tilde{\rho}^2}
\sin 2\theta +\frac{\tilde{e}}{2} \sin \theta = 0 .
\end{equation}

The determination of the extremum of the functional requires one to solve
the previous nonlinear second order differential equation with corresponding
boundary conditions for $\theta (\tilde{\rho},\tilde{e})$ at $\tilde{\rho}=0$
and $\tilde{\rho}=1$.

\begin{figure}[h]
\centering
\includegraphics[angle=0,width=0.9%
\columnwidth]{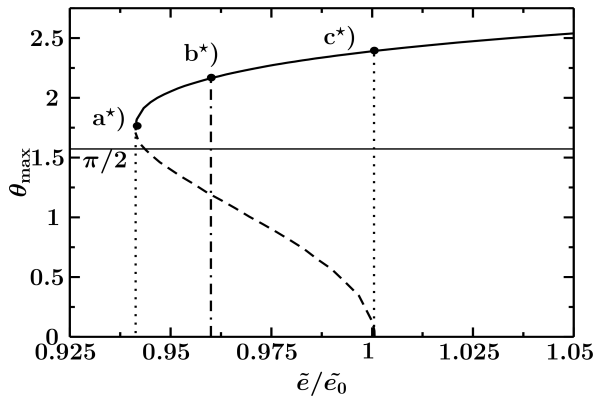}\caption{Maximun value of $\protect\theta (\tilde{\protect\rho})$ as a
function of $\tilde{e}/\tilde{e}_{0}$.
The solid and dashes line respectively corresponds to two differents vortex states
with different free energy minimum $\tilde{F}$ (Fig.~\ref{FIGE_BIS}).
Particular selected  points correspond to : a$%
^{\star }$) $\tilde{e}_{c-\mathrm{min}}\approx 0.94\tilde{e}_{0}$, b$^{\star
}$) $\tilde{e}_{c-\mathrm{abs}}\approx 0.96\tilde{e}_{0}$and  c$^{\star }$) $%
\tilde{e}_{c-\mathrm{max}}=\tilde{e}_{0}$.
Vertical dashed-dotted lines
correspond to critical fields for transition between absolute stable states.
The range of field values delimited by vertical dotted line corresponds to
the maximal extension of possible electric hysteresis phenomena.}
\label{FIGE}
\end{figure}

\begin{figure}[!h]
\centering
\includegraphics[angle=0,width=0.9%
\columnwidth]{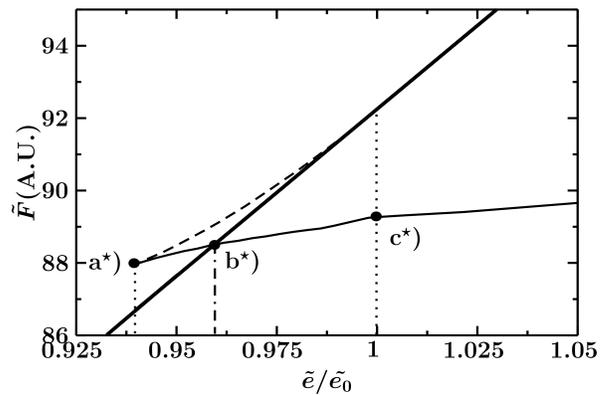}
\caption{Energy $\tilde{F}$ as a function of $\tilde{e}/\tilde{e}_{0}$ for
different polarization distribution : homogeneous up-state (thick solid
line), vortex state $\mathrm{d} \protect\theta_{\mathrm{max}}/\mathrm{d}
\tilde{e} <0 $ (dashed line) and vortex state $\mathrm{d} \protect%
\theta_{\mathrm{max}}/\mathrm{d} \tilde{e} >0 $ (solid line).
Particular selected  points correspond to : a$%
^{\star }$) $\tilde{e}_{c-\mathrm{min}}\approx 0.94\tilde{e}_{0}$, b$^{\star
}$) $\tilde{e}_{c-\mathrm{abs}}\approx 0.96\tilde{e}_{0}$and  c$^{\star }$) $%
\tilde{e}_{c-\mathrm{max}}=\tilde{e}_{0}$. }
\label{FIGE_BIS}
\end{figure}

\begin{figure}[!h]
\centering
\includegraphics[angle=0,width=0.9%
\columnwidth]{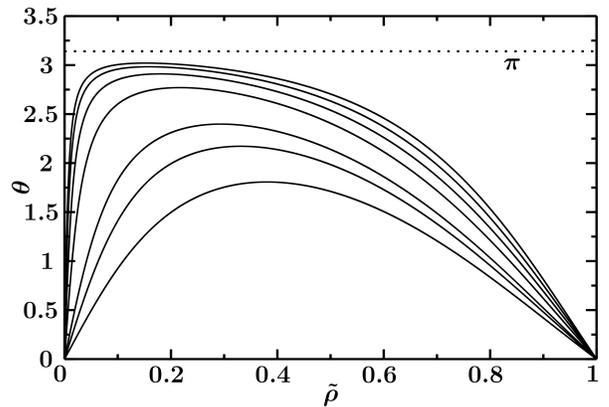}
\caption{$\protect\theta$ as a function of $\tilde{\protect\rho}$ for
different electric field values, from the bottom to the top: a$^\star$) $%
\tilde{e}_{c-\mathrm{min}} \approx 0.94 \tilde{e}_0$ , b$^\star$) $\tilde{e}%
_{c-\mathrm{abs}} \approx 0.96 \tilde{e}_0$, c$^\star$) $\tilde{e}_{c-%
\mathrm{max}} = \tilde{e}_0$,  $1.2 \tilde{e}_0$,  $1.4
\tilde{e}_0$,  $1.6 \tilde{e}_0$,  $1.8 \tilde{e}_0$. }
\label{FIGF}
\end{figure}

We obtain the
equilibrium state by using a numerical method suitable for solving the
nonlinear state equation Eq.~\ref{EQ_ETAT_NONLIN}. We have plotted in Fig.~%
\ref{FIGE} the evolution of the maximum  $\theta _{\mathrm{max}}=\max \theta
(\tilde{r})$,  as function of rescaled electric field $\tilde{e}/\tilde{e_{0}%
}$ with $\tilde{e}=\tilde{e}_{0}\approx 29.36$. There exist a critical field
$\tilde{e}_{c-\mathrm{min}}^{\star }\approx 0.94\tilde{e}_{0}$
upon which there exist non zero solutions of Eq.~(\ref{EQ_ETAT_NONLIN}).
Between $\tilde{e}_{c-\mathrm{min}}^{\star }$ and $\tilde{e}_{c-\mathrm{max}%
}^{\star }=\tilde{e}_{0}$ there are two possibilities for vortex state with
two different values for $\theta _{\mathrm{max}}$. From $\tilde{e}_{c-%
\mathrm{min}}^{\star }$ we observe two branches, the first one with negative
slope reaches $0$ at $\tilde{e}_{c-\mathrm{max}}^{\star }$ and the second one
with positive slope which approaches 
$\tilde{e}_{c-\mathrm{max}}^{\star }$.
In order to determine the state which is the most stable we have plotted in
Fig.~\ref{FIGE_BIS} the evolution of the free energy $\tilde{F}$ as a
function of $\tilde{e}/\tilde{e_{0}}$ from the different kinds of solutions.
With the dotted line, we have represented the free energy $\pi h\tilde{e}$
of the homogeneous up-oriented state. From $\tilde{e}_{c-\mathrm{min}%
}^{\star }$ we observe also the evolution of the free energy which
corresponds to the two branches previously described in Fig.~\ref{FIGE}. The
curve with dashed line in Fig.~\ref{FIGE_BIS} corresponds to the part of the
curve with dashed line in Fig.~\ref{FIGE}. In that case the energy is
greater than the energy $\pi h\tilde{e}$ of the trivial  solution $\theta =0$
all over the range of fields $\tilde{e}_{c-\mathrm{min}}^{\star }-\tilde{e}%
_{c-\mathrm{max}}^{\star }$ for which this solution exists, so that it
doesn't corresponds to the more stable state which remains the homogeneous
up state. The curve with solid line in Fig.~\ref{FIGE_BIS} corresponds to
the part of the curve with solid line in Fig.~\ref{FIGE} also exhibit
an energy greater than $\pi h\tilde{e}$ up to $\tilde{e}=\tilde{e}_{c-%
\mathrm{abs}}^{\star }\approx 0.96\tilde{e}_{0}$ so that the vortex state
less stable than the homogeneous state. On the contrary, upon reaching $\tilde{e}=%
\tilde{e}_{c-\mathrm{abs}}^{\star }$ the energy of the second solution, is
lower than $\pi h\tilde{e}$. As a consequence this second solution
corresponds to the more stable solution in the presence of an applied field
greater than $\tilde{e}_{c-\mathrm{abs}}^{\star }$. At $\tilde{e}=\tilde{e}%
_{c-\mathrm{max}}^{\star }$ we observe the existence of a single
solution which corresponds to the prolongation of the second solution 
in the range $\tilde{e}_{c-\mathrm{min}}^{\star }-\tilde{e}_{c-\mathrm{max}}^{\star }$ [Fig.~\ref{FIGE_BIS}].
We have represented in Fig.~\ref{FIGF} the distribution $\theta (\tilde{r})$
for different electric fields. The shape observed for the metastable state
at $\tilde{e}_{c-\mathrm{min}}^{\star }$ is  slighly  asymmetric.
It becomes more  asymmetric as $\tilde{e}$ increases. For the highest
electric field  the $\theta (\tilde{r})$ distribution is highly asymmetric
which reveals an important characteristic of the vortex state.

%
%

\section{DISCUSSION OF EXPERIMENTAL SITUATIONS}

The experimental study of circular closure or vortex domains began
experimentally with the study by Dawber et al. \cite{DAWBER_2003} of
nucleation and growth of ferroelectric domains in unpoled PZT films. By
measuring the frequency response of thin films to ac signals, they found
that a resonance was observed with frequency equal to the reciprocal of the
specimens perimeter. This relationship was interpreted as nucleation at the
edge, propagation around the circumference, leaving an unswitched core at
the center. The effect was not proportional to the diameter or area of the
specimen, only perimeter, as verified by using square and rectangular
samples of very different aspect ratios, from 1 $\times $ 1 $\mu $m to 130 $%
\times $ 180 $\mu $m (4 $\mu $m to 620 $\mu $m). 
However, these circular closure domains were not initially tested via direct
spatial observation, either electron microscopy, optical microscopy, or PFM.
It is important in the present context that these closure domains were
observed only in films that had not been pre-poled but which had simply been
cooled from above the Curie temperature. Even more important, this was
observed to be a low-field phenomenon. It occurred only for $E<<100\mathrm{kV/cm}$
 and in fact could occur for fields $\times $10 smaller than the
nominal coercive field $E_{c}$. For larger applied fields, the switched
polarization was homogeneous, from $+P$ to $-P$, with no unswitched hole in
the center. This early work implied two distinct coercive fields :  a low
coercive field for vortex instability, and a higher coercive field for
homogeneous switching. However, these experiments showed that the speed of
propagation of the closure domain was given ballistically \cite{STEVENSON_1966} by the formula

\begin{equation}
\tau =\frac{2\lambda }{v_{0}}\exp \left( \frac{R}{2l}\right) ,
\end{equation}%
where $R$ is the disc radius, $l$ is the phonon wavelength causing
viscuous drag with coefficient and $v_{0}$  the domain wall velocity.

\begin{equation}
D=\frac{1}{2}\rho Av_{0}^{2},
\end{equation}

$\rho $ is the sample density and $A$ is the cross section.
Velocity $v_{0}$ was fitted to be $3\mathrm{m/s}$ in PZT.
In the present model we not only assume a much faster wall velocity, but we
neglect damping.

The second experimental paper \cite{GRUVERMAN_2008} revealed spatially (via
PFM) and with 100 ns time resolution, closure domains in circular discs at
higher fields (ca. $70 \mathrm{kV/cm}$) but not in squares of the same
diameter (2 $\mu$m) and thickness, indicating a significant role of boundary
conditions. A model simulation, based upon ferromagnetic
Landau-Lifshitz-Gilbert equations reproduced the experiments well and also
indicated that this is a low-field phenomenon. Further modeling by the
authors of Ref.~\cite{GRUVERMAN_2008} showed that there was a critical size
for these closure domains, which for discs was ca. 2-micron diameter and
smaller for squares. Triangular specimens in the simulation did not exhibit
closure domains for any field or lateral size.

\begin{figure}[!h]
\centering
\includegraphics[angle=0,width=0.5\columnwidth]{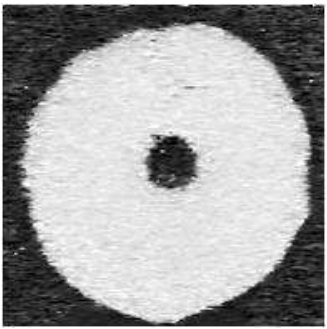}
\caption{Closure domain with circular unswitched center (adapted from Ref.~\cite{GRUVERMAN_2008})}
\label{FIGJIM}
\end{figure}

The third experimental paper\cite{SCOTT_2008} showed that the circular
unswitched center[Fig.~\ref{FIGJIM}] in the closure domains was only
metastable and disappeared in zero field after ca. 24 hours. At longer times
the outer edge became faceted along high-symmetry planes, indicating
relaxation caused by crystalline anisotropy\cite{GANPULE_2002,LUKYANCHUK_2014}
 
Most recently the study of closure domains in nano-dots\cite{SCHILLING_2009}
revealed that they arise from experimentally cooling through $T_{c}$ with
very small or zero field. This time simulation was done 
via a standard finite element micromagnetics software MAGPAR \cite{SCHOLZ_2003} and again
supported the experiments. The MAGPAR calculation had been done as a simple
magnetic domain analogy and surprisingly displays the basic qualitative
features of the experiment, although unlike the present work it offers no
explanation for the real mechanism or numertical values. It does illustrate
the sensitivity to boundary conditions and size in the problem, and the
basic Bessel-function-like solutions.

Taken as a whole, these experiments show, in agreement with the theory
above, that there are two coercive fields in ferroelectric nano-dots: A very
small coercive field for vortex instability switching $E^{\mathrm{vortex}}_c$%
; and a large coercive field for conventional switching $E_c$ from
homogeneous $+P$ to $-P$ states. There is an indication that there may be a
critical radius for these effects, but that is uncertain; none was found in
Ref.\cite{DAWBER_2003}.

These closure domains may have important application for memory storage in
ferroelectric memories (FRAMs): as Prosandeev et al. have shown,\cite%
{PROSANDEEV_2007,PROSANDEEV_2008} these configurations permit very high
bit-density and can be reversed by application of external electric fields.

\section{DYNAMICAL PROPERTIES}

Up to now we have studied the static case and determined the equilibrium
polarization state in the presence of an electric field. We now turn our
attention to discuss the essence of the dynamical polarization reversal
process on the basis of the Landau-Khalatnikov equation of motion. The
evolution of the polarization with time is expressed as follow :

\begin{equation}  \label{LKeq}
\frac{\partial P_{i}}{\partial t}=-L_{i}\frac{\delta F}{\delta P_{i}},
\end{equation}%
where $L_{i}$ are the damping coefficients, $i$ =$\left\{ \varphi ,z\right\} $.


In our model the time evolution of the polarization is parametrized by the inclination angle $\theta $
and corresponding  Landau-Khalatnikov looks like

\begin{equation}
\frac{\partial \theta }{\partial t} = - L \frac{\delta F}{\delta \theta}.
\end{equation}

The switching mechanism is initiated by a weak angle instability $\delta
\theta $. For a field $e>\tilde{e}_{c}$ we observe the polarization
switching by mean of progressive change of the angle $\theta $. This
phenomena is inhomogeneous and first affects a region located close to the
half-radius of the electrode. Then we distinguish two stages, the first one
which is almost symmetric with respect to the half-radius and a second one
which consists in a radial extension of the switched region.

\begin{figure}[!h]
\centering
\includegraphics[angle=0,width=0.9%
\columnwidth]{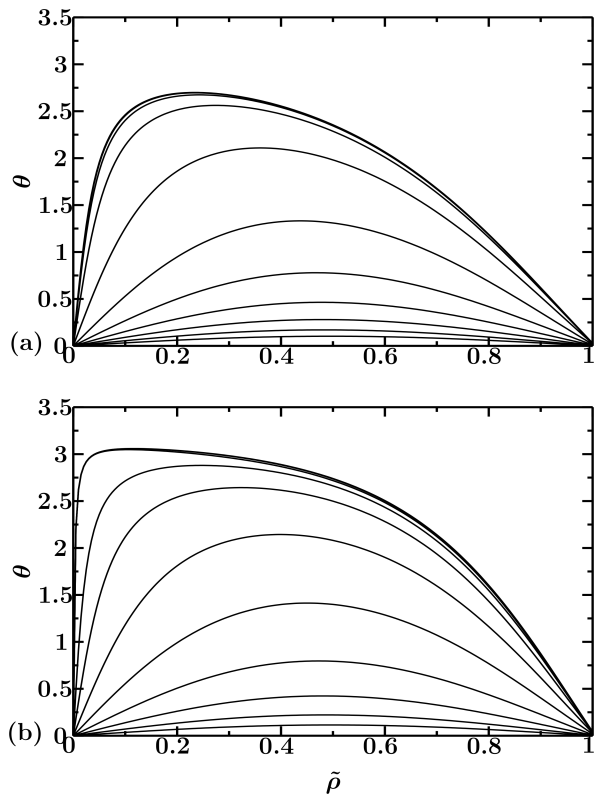}
\caption{From the bottom to the top : evolution of the distribution $\protect%
\theta(\tilde{\protect\rho})$ with time for $\tilde{e}=1.2\tilde{e}%
^{\star}_{c-\mathrm{min}}$ (a) and $\tilde{e}=2\tilde{e}^{\star}_{c-\mathrm{%
min}}$ (b). At the initial time the polarization is homogeneous and up
oriented $(\protect\theta = 0)$. The time-step values and the value of $L$
are chosen to favor the visualization of the evolution of $\protect\theta(%
\tilde{\protect\rho})$ until the final state is reached (curve at the top of
the figure). }
\label{FIGT}
\end{figure}

We present in Figs.~\ref{FIGT}(a) and ~\ref{FIGT}(b) the time evolution of
the polarization profile during the switching process. On the whole this
polarization reversal exhibits characteristics similar to the process in
thin films described in Fig.~\ref{FIGA} : the switching occurs in the orthoradial direction 
and  is followed by a growth in a perpendicular radial direction.
This suggests that behavior of polarization
component could be an universal characteristic of the switching mechanism in
ferroelectric materials.

\begin{figure}[!h]
\centering
\includegraphics[angle=0,width=0.7\columnwidth]{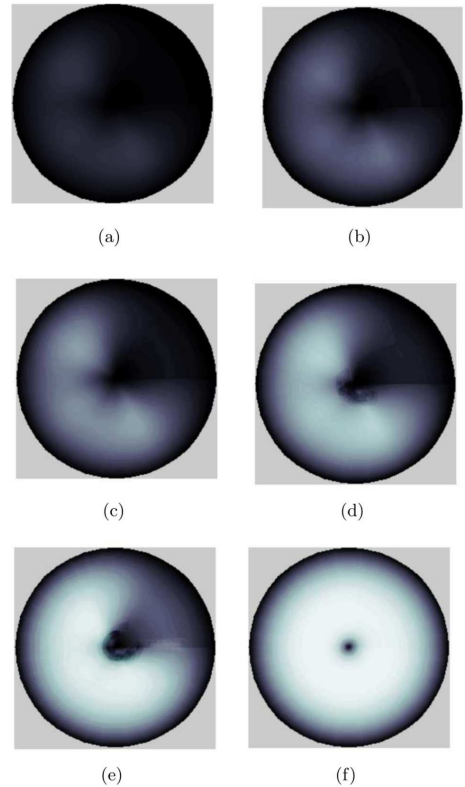}
\caption{Evolution of the polarization texture below the circular electrode
during the switching process. With black color : the "up" state, with white
color the "down" state.}
\label{FIG2D}
\end{figure}

In order to test the capability of our model to describe the experimental
behavior, we have carefully examined the results reported by Gruverman\cite%
{GRUVERMAN_2008}, and compared with those given by our model by mean of
quasi-2D numerical simulation. Main characteristics of PFM measurements are
: i) nucleation of down state at different places, ii) orthoradial rotation,
iii) radial rotation which leads to doughnut with small central peripheral
unswitched region. Our simulation considers that there exist latent nuclei
which corresponds to regions with instabilities $\delta \theta $, at
different places that  are  sample-dependent and adopted Gruverman's data.
Fig.~\ref{FIG2D} shows the two-dimensional angular distribution of $\theta (\tilde{\rho},\varphi )$.
The initial state and a stage which corresponds to the appearance of many switched regions are
respectively reproduced in Fig.~\ref{FIG2D}(a) and (b).
The next stages reproduce orthoradial extension [Fig.~\ref{FIG2D}(c) and (d)] which can be
interpreted as a time delay between polarization rotation induced by the inhomogeneous
nature of the instabilities distribution $\delta \theta (\tilde{\rho}%
,\varphi )$. It leads to doughnut shape with large central unswitched
region. The last stage corresponds the radial rotation which contributes to
reduction of the central region area [Fig.~\ref{FIG2D}(e)]. At the end of the
process when the stable state is reached [Fig.~\ref{FIG2D}(f)], the doughnut
shape reported by Gruverman,  with small central region and peripheral
unswitched region is obtained.

\section{CONCLUSION}

We have considered a finite ferroelectric  cylinder with top and bottom
electrodes used to apply an electric field. By using a thermodynamical approach we
have demonstrated that there exist non-trivial mathematical solutions
induced by the boundary conditions at the perimeter of the cylinder. These solutions  correspond to a
metastable polarization state and are observed only when an external
electric field is greater than a coercive value $\tilde{e}_{\mathrm{coercive}%
}$. We have also demonstrated that the switching mechanism have the
behaviour of a field-induced discontinous transition with hysteresis in the
interval $\tilde{e}_{c-\mathrm{min}}^{\star }-\tilde{e}_{c-\mathrm{max}%
}^{\star }$. However, the kinetic contribution can influence the precise
value of coercive field  $\tilde{e}_{\mathrm{coercive}}$, which would not strongly deviate from the range 
$\tilde{e}_{c-\mathrm{min}}^{\star}-\tilde{e}_{c-\mathrm{max}}^{\star }$.
 One can argue that the order of magnitude of $\tilde{e}_{\mathrm{coercive}}
$ is given by the value $\tilde{e}_{0}$ itself and is governed by the
boundary condition at the perimeter of the electrode.
One can estimate the coercive field  for the vortex-formationg switching by 
the field  $E_{\mathrm{c}}^{\mathrm{vortex}}=\alpha _{0}\left( \frac{\zeta _{0}}{R}\right) ^{2}P\tilde{e}_{0}$ and compare it with the classical thermodynamic coercive field $E_{%
\mathrm{c}}$. These two fields are found to be equal when $R=R_{0}\approx
8.73\zeta _{0}$. Usually $R>R_{0}$ and $E_{\mathrm{c}}^{\mathrm{vortex}}<E_{%
\mathrm{c}}$.

In the case of Gruverman's \cite{GRUVERMAN_2008} study, one can estimate
from the experimental parameters the value of $E_{\mathrm{c}}^{\mathrm{vortex%
}}$. It is found to be many times lower than the rectilinear-domain coercive
field. In these conditions it seems reasonable that the existence of a
coercive field for vortex formation has not been reported in Gruverman's
\cite{GRUVERMAN_2008} article. The evolution of the polarization texture in
two-dimensional films allowed us to describe the vortex formation. The
calculated final steady state is similar the unusual domain pattern with
doughnut shape previously experimentally observed only in the case of
circular capacitor \cite{GRUVERMAN_2008}. The metastable characteristic of
the doughnut distribution \cite{SCOTT_2008} is also followed from our consideration
since it vanishes when the electric field is suppressed.
The fact that the central
and peripheral unswitched regions are very small is also consistent with the
results given by the present theory which predicts polarization saturation
with $\theta =\pi $ along the radius and unswitched regions with $\theta =0
$  at the center and at the peripheral edge of the electrode. The effect of
the electric field tends to enlarge the proportion of the region where the
polarization saturates and to decrease the central and the peripheral
regions. The dimensions of the unswitched regions are very small, which is
revelant for an applied electric field much higher that the value given by
our calculation. Hence one should carefully investigate experimentally the
role of the applied field on the possible vortex formation.
 Since the polarization reversal rotationnal mechanism  costs less energy than the
classical one provided by change  in the modulus, usually observed in thin films, one can
expect that this alternative switching mechanism, induced by both the
geometry and the interface properties would be especially useful in memory-storage.

\begin{acknowledgments}
We thank  G. Catalan,  M. Gregg, A. Schilling and A. Gruverman for
stimulating exchange in frame of the France-UK
collaboration program Alliance.
\end{acknowledgments}

\bibliographystyle{./BIBLIOGRAPHIE/apsrev4-1.bst}

\end{document}